\documentclass[12pt]{article}
\usepackage{ucs}
\usepackage[utf8x]{inputenc}
\usepackage{braket}
\usepackage{dsfont}
\usepackage{xcolor}
\usepackage{cite} %to produce [1-3] instead of [1,2,3]
\usepackage[linktocpage]{hyperref}
\usepackage{amscd,amsmath,amssymb}
\usepackage{graphics,graphicx}
\usepackage[mathscr]{euscript}
\numberwithin{equation}{section} %%
\def\a{\alpha}
\def\b{\beta}

\def\d{\delta}
\def\e{\epsilon}

\def\m{\mu}
\def\n{\nu}

\def\r{\rho}
\def\s{\sigma}

\def\ve{\varepsilon}

\def\be{\begin{equation}}
\def\ee{\end{equation}}
\def\gammarr{\begin{array}{rll}}
\def\ea{\end{array}}
\def\bea{\begin{eqnarray}}
\def\eea{\end{eqnarray}}
\def\N2{$N{=}2$}

\def\>{\rangle}
\def\<{\langle}
\def\+{\dagger}
\def\={\ =\ }

\begin{document}

\title{
\hfill\\
\hfill\\
Spontaneously Broken $3d$ Hietarinta/Maxwell Chern-Simons Theory and Minimal Massive Gravity
\\[0.5cm]
}

\author{
Dmitry Chernyavsky$\,{}^{a}$, Nihat Sadik Deger$\,{}^{b,c}$ and Dmitri Sorokin$\,{}^{d,e}$
}

\date{}

\maketitle
\vspace{-1.5cm}

\begin{center}
\vspace{0.5cm}\textit{\small
${}^a$ School of Physics, Tomsk Polytechnic University,\\
634050 Tomsk, Lenin Ave. 30, Russia	}

\vspace{0.5cm}\textit{\small
${}^b$ Department of Mathematics, Bogazici University,\\
Bebek, 34342, Istanbul, Turkey \\
\centerline {$\&$}
${}^c$ Max-Planck-Institut f\"ur Gravitationsphysik
(Albert-Einstein-Institut) \\
Am M\"uhlenberg 1, D-14476 Potsdam, Germany
}

\vspace{0.5cm}
\textit{\small
${}^d$ I.N.F.N., Sezione di Padova \\
\centerline {$\&$}
${}^e$ Dipartimento di Fisica e Astronomia ``Galileo Galilei",  Universit\`a degli Studi di Padova, \\
Via F. Marzolo 8, 35131 Padova, Italy
}
\end{center}
\vspace{5pt}

\abstract{We show that  minimal massive $3d$ gravity (MMG) of \cite{Bergshoeff:2014pca}, as well as the topological massive gravity, are particular cases of a more general `minimal massive gravity' theory (with a single massive propagating mode) arising upon spontaneous breaking of a local symmetry in a Chern-Simons gravity based on a Hietarinta or Maxwell algebra. Similar to the MMG case, the requirements that the propagating massive mode is neither tachyon nor ghost and that the central charges of an asymptotic algebra associated with a boundary CFT are positive, impose restrictions on the range of the parameters of the theory.}

\noindent

\thispagestyle{empty}

%\vfill
%\vskip 5.mm
%\hrule width 5.cm
%\vskip 2.mm
%{\scriptsize
%\noindent e-mails: {\tt sorokin@pd.infn.it
%}}

\newpage

\setcounter{footnote}{0}

\tableofcontents

%\newpage
\section{Introduction}

Three-dimensional gravity theories have attracted great deal of attention since the early 80s as simpler tools for studying features of General Relativity in higher dimensions, its possible consistent modifications and extensions to quantum gravity. Since then a variety of different $3d$ gravity models with interesting geometric and physical properties have been constructed and analyzed. Among these is the minimal massive $3d$ gravity  (MMG) \cite{Bergshoeff:2014pca} which will be the focus of our attention in this paper.
This gravity model is a particular case of a class of Chern-Simons-like theories \cite{Hohm:2012vh,Bergshoeff:2014bia,Merbis:2014vja}. In contrast to the genuine $3d$ Chern-Simons theories which do not have local degrees of freedom in the bulk, the Chern-Simons-like gravities have propagating massive spin-2 modes coupled to a number of other spin-2 fields.\footnote{By ``spin-2 fields" we somewhat loosely mean $3d$ Lorentz-vector-valued one-form fields $a^{r a}=dx^\mu a^{r a}_\mu(x)$ ($r=1,2,\ldots,N$) which include a dreibein $e^a(x)$ and a dualized spin connection $\omega^a=\frac 12 \varepsilon^{abc}\omega_{bc}$.}

One of the main motivations for constructing modifications of $4d$ General Relativity which include massive gravitons is to try to explain in this way the nature of dark matter and dark energy. Three-dimensional massive gravities serve as useful toy models for studying peculiar features and issues of these theories regarding e.g. the absence of Ostrogradski ghosts etc.
An open fundamental question regarding gravity theories with massive gravitons is whether a spin-2 field mass can be attributed to spontaneous breaking of a space-time symmetry which in general can be an extension of the Poincarè group. To answer this question one should first individualize such a symmetry and then, ideally, find a mechanism generating mass of a corresponding spin-2 field similar to that of Englert–Brout–Higgs–Guralnik–Hagen–Kibble. By now, such a mechanism is not known for gauge spin-2 fields. In this situation one can resort to old constructions called Phenomenological Lagrangians (see e.g. \cite{Coleman:1969sm,Callan:1969sn,Volkov:1973vd}) which have proved useful for understanding the most general structure of symmetry breaking terms with the use of Goldstone fields on which these symmetries are realized non-linearly. A notable example is the first construction of the supergravity action with non-linearly realized local supersymmetry \cite{Volkov:1973jd} (see \cite{Bandos:2015xnf} for a review and further developments).

In this paper we would like to address the above question for $3d$ Chern-Simons-like MMG of \cite{Bergshoeff:2014pca} and, in particular,  to understand whether the presence of a massive spin-2 mode therein can be seen as an effect of (partial) spontaneous breaking of a local symmetry containing the $3d$ Poincarè group as a subgroup. We will show that this is indeed the case.\footnote{The broken symmetry under consideration is not a $3d$ Weyl symmetry which was assumed to be a source of the graviton mass in \cite{Dereli:2019bom}.}

The MMG contains three `spin-2' fields, the dreibein $e^a$, the spin connection $\omega^a$ and an additional one-form field ${h}^a$. The first two are associated with gauge fields of the local $3d$ Poincarè group generated by the translations $P_a$ and Lorentz rotations $J_a$. We would also like to treat ${h}^a$ as a gauge field associated with an additional vector generator $Z_a$ that extends the Poincarè group to a larger symmetry which is however broken in the MMG action. We will restore this larger symmetry by coupling the gauge fields $e^a$ and ${h}^a$ to a St\"uckelberg-like spin-1 Goldstone field associated with spontaneous breaking of $Z_a$-symmetry. The symmetry algebra in question is the simplest among  algebras constructed by Hietarinta \cite{Hietarinta:1975fu}, a class of finite-dimensional supersymmetry-like algebras containing higher-spin generators.\footnote{The most studied example of the Hietarinta algebras is the one in which the spin-1/2 generators of a supersymmetry algebra are replaced by their spin-3/2 counterparts. This algebra underlies the so-called Hypergravity put forward in $D=2+1$ by Aragone and Deser \cite{Aragone:1983sz} (see e.g. \cite{Zinoviev:2014sza,Bunster:2014fca,Fuentealba:2015jma,Fuentealba:2015wza,Henneaux:2015tar,Bansal:2018qyz,Rahman:2019mra,Fuentealba:2019bgb} for further studies of this theory).} The  commutators of the generators of this algebra are
\bea\label{Algebra_Maxwell}
&&
[J^a, J^b]=\e^{abc}J_c, \qquad [J^a, P^b]=\e^{abc}P_c, \qquad  [J^a, Z^b]=\e^{abc}Z_c,
\nonumber\\[2pt]
&&
[P_a,P_b]=0, \qquad [Z^a, Z^b]=\e^{abc}P_c.
\eea
Note that the commutator of $Z_a$ closes on translations, somewhat similar to supersymmetry. Notice also that this algebra is isomorphic (dual) to the three-dimensional Maxwell algebra \cite{Bacry:1970ye,Schrader:1972zd} in which the role of the generators $P_a$ and $Z_a$ gets interchanged $(P_a\leftrightarrow Z_a)$, namely
\be\label{Maxwell}
[Z_a,Z_b]=0, \qquad [P^a, P^b]=\e^{abc}Z_c.
\ee

The Chern-Simons action for gravity with the local symmetry generated by the $3d$ Maxwell algebra was constructed and studied in \cite{Salgado:2014jka,Hoseinzadeh:2014bla,Aviles:2018jzw,Concha:2019lhn}.\footnote{ Higher-spin extensions of the Maxwell algebra and corresponding gravity models were considered in \cite{Caroca:2017izc}.  See also \cite{Salgado-Rebolledo:2019kft,Caroca:2019dds} for a detailed study of the $3d$ Maxwell group, its infinite-dimensional extensions, applications and additional references.}, while its Hietarinta counterpart was considered in \cite{Bansal:2018qyz,Chernyavsky:2019hyp}.
Since from the algebraic point of view the construction of the action is the same for \eqref{Algebra_Maxwell} and \eqref{Maxwell} and the only difference between the two is the choice of the physical interpretation of the generators and corresponding gauge fields, in what follows we will call the general model under consideration the Hietarinta/Maxwell Chern-Simons  Gravity (HMCSG).

In Sections 2 and 3 we will show that augmenting the HMCSG action with terms that break linearly realized symmetry \eqref{Algebra_Maxwell} along $Z_a$ one gets an extension of the Minimal Massive Gravity. It has, in general, two more coupling terms in comparison with the MMG, but still has a single massive propagating degree of freedom,  as we show by performing the Hamiltonian analysis in Section \ref{Ha} and studying linear perturbations of the fields around an $AdS_3$ background in Section \ref{Lp}.
In Section 4, as a side remark, we demonstrate that when the parameters of the HMCSG  are restricted by a certain condition which makes its equations of motion integrable, the model reduces to a pure
Chern-Simons theory with the gauge group $SL(2,R)\times SL(2,R)\times SL(2,R)$.
In Section 6 we compute the central charges of an asymptotic symmetry algebra of the HMCSG with $AdS_3$ boundary conditions.
As in the MMG case, the requirements that the propagating massive mode is neither tachyon nor ghost and that the boundary CFT central charges are positive impose restrictions on the range of the parameters of the HMCSG theory. We analyze these restrictions for some particular cases for which the parameters of the HMCSG differ from those of the original MMG in Section 7, and conclude with comments and an outlook in Section 8.

\section{Hietarinta/Maxwell Chern-Simons gravity and its minimal massive extension}\label{Maxwell and MMG}
Let us start by reviewing the construction of a gravity action which enjoys local symmetry transformations associated to the algebra (\ref{Algebra_Maxwell}).
The algebra (\ref{Algebra_Maxwell}) has the following invariant bilinear form
\be\label{Bilinear Forms}
\langle J_a, Z_b\rangle = {\tt a}\eta_{ab},\qquad  \langle J_a, P_b\rangle =  \langle Z_a, Z_b\rangle = -\sigma\,{m} \eta_{ab},\qquad \langle J_a, J_b\rangle = \eta_{ab},
\ee
where $m$ is a parameter of the dimension of mass, $\tt a$ has the dimension of $m^{\frac 12}$, while  ${(-\sigma)}$ is an arbitrary dimensionless constant\footnote{The minus sign in front of $\sigma$ was chosen to make our convention closer to that of \cite{Bergshoeff:2014pca}. We will also set the value of the gravitational constant as $16\pi G=1$.}. The dimensions of the coefficients reflect the canonical dimensions of $[J_a]=m^0$, $[P_a]=m$ and $[Z_a]=m^{\frac 12}$.

In the case of the Maxwell algebra \eqref{Maxwell} the dimension of $Z_a$ changes to $m^2$ and the corresponding bilinear form is
\be\label{Bilinear Forms M}
\langle J_a, P_b\rangle = {\tt a}\, m\eta_{ab},\qquad  \langle J_a, Z_b\rangle =  \langle P_a, P_b\rangle = -\sigma {m^2} \eta_{ab},\qquad \langle J_a, J_b\rangle = \eta_{ab},
\ee
where now the parameter $\tt a$ is dimensionless.

The bilinear form \eqref{Bilinear Forms} is used to construct the Chern-Simons action (in which the wedge product of the differential forms is implicit)
\be\label{Action_CS}
S=\frac 1{2 m}\int_{\mathcal{M}_3}\langle \mathbf{A}d\mathbf{A}+\frac{2}{3}\mathbf{A}^3\rangle \, ,
\ee
for the  gauge field one-form $\mathbf{A}$ taking values in the algebra (\ref{Algebra_Maxwell})
\be\label{bfA}
\mathbf{A}=e^a P_a+\omega^a J_a+{h}^a Z_a.
\ee
Explicitly, for the components of \eqref{bfA} the  action \eqref{Action_CS} takes the following form
\be\label{Action_Gravity0}
S_{HCS}=\frac{1}{2}\int_{\mathcal{M}_3} \,\left[  \frac{2\tt a}m\,{h}^a R_a-\sigma (2e^a R_a+{h}^{a}\nabla {h}_{a})+\frac 1 { m}\left(\omega^a d\omega_a+\frac 13\varepsilon_{abc}\omega^{a}\omega^b\omega^c\right)\right],
\ee
where
\be
\nabla {h}^a=d{h}^a+\ve^{abc}\omega_b {h}_c, \qquad R^a=d\omega^a+\frac12 \ve^{abc}\omega_b\omega_c.
\ee
The Hietarinta Chern-Simons (HCS) action \eqref{Action_Gravity0} is invariant (up to a boundary term) under the infinitesimal gauge transformations
\bea\label{Transformations_Gauge}
&&
\d e^a=\nabla \ve^a_P+\varepsilon^{abc}({h}^b \ve^c_Z+e^b \ve^c_J),
\nonumber\\[2pt]
&&
\d {h}^a=\nabla \ve^a_Z+\varepsilon^{abc}{h}^b \ve^c_J,
\nonumber\\[2pt]
&&
\d \omega^a=\nabla \ve_J^a.
\eea
Note that the term $\frac{2\tt a}m {h}^a R_a$ in \eqref{Action_Gravity0} can be absorbed by the term $2e^a R_a$ upon the field redefinition $e^a \rightarrow e^a-\frac {\tt a}{\sigma m} {h}^a$. So, without  loss of generality, instead of \eqref{Action_Gravity0} we will deal with the action
\be\label{Action_Gravity}
S_{HCS}=\frac 12\int_{\mathcal{M}_3} \,\left[{-\sigma} (2e^a R_a+{h}^{a}\nabla {h}_{a})+\frac 1 { m}\left(\omega^a d\omega_a+\frac 13\varepsilon_{abc}\omega^{a}\omega^b\omega^c\right)\right].
\ee
Also note that if instead of the Hietarinta algebra \eqref{Algebra_Maxwell}, we had used the Maxwell algebra \eqref{Maxwell} and the corresponding bilinear form \eqref{Bilinear Forms M} as the basis for constructing the action \eqref{Action_CS}, instead of \eqref{Action_Gravity} we would get
\be\label{Action_Gravity1}
S_{MCS}= \frac 12\int_{\mathcal{M}_3} \,\left[2 {\tt a} e^a R_a-m\,{\sigma} (2{h}^a R_a+e^{a}\nabla e_{a})+\frac 1 { m}\left(\omega^a d\omega_a+\frac 13\varepsilon_{abc}\omega^{a}\omega^b\omega^c\right)\right].
\ee
In this action the role of the dreibein $e^a$ (associated with the Poincarè translations) and of the additional spin-2 field ${h}^a$ get interchanged in comparison to  \eqref{Action_Gravity}. \footnote{Notice that in the Maxwell case the dimension of ${h}_a$ gets changed in comparison with the Hietarinta case in accordance with the change of the dimension of $Z_a$ in \eqref{Bilinear Forms M}.} Now we can absorb the first term of \eqref{Action_Gravity1} into its second term by redefining ${h}^a\to {h}^a-\frac{\tt a} {\sigma m}e^a$ and get
\be\label{Action_Gravity1-1}
S_{MCS}= \frac 12\int_{\mathcal{M}_3} \,\left[-{m}\, \sigma\, (2{h}^a R_a+e^{a}\nabla e_{a})+\frac 1 { m}\left(\omega^a d\omega_a+\frac 13\varepsilon_{abc}\omega^{a}\omega^b\omega^c\right)\right]\,.
\ee
So, if one insists on associating the genuine graviton field with the Poincar\'e generator $P_a$, one concludes that the Maxwell Chern-Simons (MCS) gravity based on \eqref{Action_Gravity1-1}  actually does not have the standard Einstein term $e^aR_a$. In this respect the Maxwell Chern-Simons gravity \eqref{Action_Gravity1-1} can be regarded as a deformation  of the ``exotic" Einstein gravity considered e.g. in \cite{Witten:1988hc}. The parity-odd first order action of the latter is obtained from \eqref{Action_Gravity1-1} by removing its first term.

From the Chern-Simons structure of the action \eqref{Action_CS} it follows that the models under consideration do not have propagating degrees of freedom in the $3d$ bulk.

\subsection{Breaking the Hietarinta symmetry}
We would now like to generate non-trivial bulk dynamics (and mass) of fields  in the above Hietarinta/Maxwell Chern-Simons model by adding to the action \eqref{Action_Gravity} terms which can be associated with a spontaneous breaking of the Hietarinta symmetry \eqref{Algebra_Maxwell} down to its Poincar\'e subalgebra.

\subsubsection{Spontaneous breaking of the rigid symmetry}
By the Goldstone's theorem, the spontaneous breaking of a rigid (global) continuous symmetry
is characterized by the appearance of massless Nambu-Goldstone fields associated with broken symmetry generators. In the case under consideration these are the vector generators $Z_a$ and the corresponding Goldstone field is a vector field $A_a(x)$ of mass dimension $m$ \cite{Bansal:2018qyz} which should not be confused with the Chern-Simons one-form \eqref{bfA}. The Goldstone vector field appears in the Cartan one-form \footnote{For the details of the model see \cite{Bansal:2018qyz} which in turn is based on the Volkov-Akulov construction \cite{Volkov:1972jx,Volkov:1973ix} of Lagrangians with spontaneously broken and non-linearly realized supersymmetry.}
\bea\label{CF}
\Omega_0&=&g_0^{-1}dg_0=E_0^aP_a+{H}_0^aZ_a,\nonumber\\
E_0^a&=&dx^a-\frac {f^{-2}}2 \varepsilon^{abc}A_bdA_c,\\
{H}_0^a&=&f^{-1}dA^a(x),\nonumber
\eea
where
\be\label{g0}
g_0=e^{x^a P_a}e^{f^{-1}A^a(x)Z_a} \, ,
\ee
is a Hietarinta group element with $x^a$ being a flat $3d$ space-time coordinate and $f$ being a symmetry breaking parameter  of mass-dimension $m^{\frac 32}$. The subscript 0 indicates that, at this moment, we are dealing with a rigid symmetry with respect to which the one-form \eqref{CF} is invariant under the transformation
\be\label{rigid}
g_0 \to e^{\ve^a_JJ_a}e^{\ve^a_PP_a} e^{\ve^a_ZZ_a}g_0 \, ,
\ee
where the parameters are $x$-independent. The spontaneously broken symmetry associated with $\ve^a_ZZ_a$ is realized on $x^a$ and the Goldstone field $A_a(x)$ infinitesimally as a non-linear transformation  \cite{Bansal:2018qyz}
\be\label{nl}
 \delta x^a= \frac {f^{-1}}2 \,\varepsilon^{abc}\ve_{b\, Z}A_c (x), \qquad \delta A^a=f\ve^a_Z-\frac {f^{-1}}2\,\varepsilon_{dbc}\ve_{Z}^{b}\,A^c\,\partial^dA^a.
\ee
The unique Lagrangian for $A_a(x)$ with the minimal number of derivatives (up to two) which is invariant under \eqref{nl} is of the Volkov-Akulov type and has the following form
\bea\label{S1}
S_{1}&=& \frac{\mu_1 f^{2}}{3!}\int\,\varepsilon_{abc} E^a_0E^b_0E^c_0\\
&=&\mu_1\int\,d^3x\left(f^2+ \frac 12 \varepsilon^{abc}A_a\,\partial_bA_c-\frac{f^{-2}}{8}\,\varepsilon^{abc}\varepsilon^{def}\,A_a\,A_d\,\partial_eA_b\,\partial_fA_c\right),\nonumber
\eea
where $\mu_1$ is a dimensionless constant parameter.

Note that a would be third-order derivative term in \eqref{S1} vanishes.
Interestingly, the action \eqref{S1} contains the Abelian Chern-Simons term for  $A_a$, while the presence of the quartic term breaks $U(1)$ gauge invariance of the CS action and makes propagating a scalar mode of $A_a$ which happens to be of a Galileon type (see \cite{Bansal:2018qyz} for details).
Therefore, the spontaneous breaking of the Hietarinta symmetry produces the vector Goldstone field which has only one dynamical degree of freedom.

Using the components of the Cartan form \eqref{CF} one can also construct a Hietarinta-invariant term which is of the third order in derivatives of $A_a(x)$
\be\label{3rd}
S_2=\mu_2 f^{\frac 53}\int \varepsilon_{abc}{H}_0^aE^b_0E^c_0,
\ee
where $\mu_2$ is a dimensionless parameter.
Modulo  total derivatives, it has the following explicit form
\be\nonumber
S_2=-\frac{\mu_2 f^{-\frac {10}3}}8\int \varepsilon_{abc}  dA^adA^bdA^c\,A^2.
\ee
Also note that two more possible contributions to the Goldstone field action are actually total derivatives
\bea\label{3L}
S_{3,4}&=&\int \varepsilon_{abc}\left(\mu_3f^{\frac 43}{H}_0^a{H}_0^bE^c_0+\mu_4f{H}_0^a{H}_0^b{H}_0^c\right)\\
&=& \int \varepsilon_{abc}\left(\mu_3f^{-\frac 23}d(A^adA^bE^c_0)+\mu_4f^{-2}dA^adA^bdA^c\right)\,.\nonumber
\eea
To recapitulate, the actions \eqref{S1}-\eqref{3L} are manifestly invariant under Lorentz rotations, Poincar\'e translations and rigid Hietarinta symmetry \eqref{nl}. The last one acts as a (non-linear) shift on the Goldstone field $A_a$ and thus is spontaneously broken by the vacuum solution $A_a=0$.
\subsubsection{Gauging the non-linearly realized symmetry}
To couple the Goldstone field $A_a(x)$ to the gauge fields \eqref{bfA}, we should covariantize the Cartan form \eqref{CF} which makes it invariant under the transformation \eqref{rigid} whose parameters are promoted to functions of the space-time coordinates $x^\mu$. The result is
\be
\Omega= g^{-1}(d+\mathbf{A})g=E^a P_a+\omega^a J_a + {H}^a Z_a,
\ee
where now
 \be
 g=e^{\phi^a(x) P_a} e^{f^{-1}A^a(x) Z_a} \, ,
 \label{element}
 \ee
with $\phi^a(x)$ being an arbitrary $3d$ vector function and
\bea\label{One-forms}
&&
E^a=e^a+\nabla\phi^a+f^{-1}\varepsilon^{abc} {h}_b A_c-\frac{f^{-2}}2 \varepsilon^{abc} A_b \nabla A_c,
\nonumber\\[2pt]
&&
{H}^a={h}^a+f^{-1}\nabla A^a.
\eea
The gauge group acts on $\phi^a$ and $A^a$ as follows
\bea\label{dfA}
&&
\d \phi^a = -\ve_P^a-\varepsilon^{abc}(\ve_{Zb}{h}_c+\ve_{Jb}\phi_c),
\nonumber\\[2pt]
&&
\d A^a=-f\ve^a_Z-\varepsilon^{abc}\ve_{Jb} A_c.
\eea
Combined with the variations of the gauge fields \eqref{Transformations_Gauge}, the action of the gauge transformations  on \eqref{One-forms} reduces to their Lorentz rotations
\be\label{Transformations_Gauge_Matter}
\d_J E^a=-\ve^{abc}\ve_{Jb} E_c, \qquad \d_J{H}^a=-\ve^{abc}\ve_{Jb} {H}_c,
\ee
leaving the one-forms \eqref{One-forms}  invariant under the action of the transformations along  $P_a$ and $Z_a$.

The following comment is now in order. The vector $\phi^a(x)$ might be thought of as a Goldstone (St\"uckelberg) field associated with breaking of the local Poincar\'e translations. However, this ``breaking" does not result in changing the number of the physical (on-shell) degrees of freedom of the dreibein $e^a$. The reason is that, in addition to the invariance under local Hietarinta symmetry, the Chern-Simons gravity action \eqref{Action_Gravity0} is invariant under the $3d$ diffeomorphisms
\be\label{dif}
x^\mu \to x^\mu+\zeta^\mu(x).
\ee
Under the diffeomorphisms the dreibein transforms as follows
\be\label{dedif}
\delta e^a =\nabla (\xi^\mu e^a_\mu)-\ve^{abc}(\xi^\mu\omega_{\mu b}) e_c+i_\xi\nabla e^a.
\ee
Comparing \eqref{dedif} with \eqref{Transformations_Gauge} we see that the first and the second term in \eqref{dedif} can be associated, respectively, with local Poincar\'e translations and Lorentz rotations. Regarding the third term, since on the mass shell   $\nabla e^a=-\varepsilon^{abc}{h}_b{h}_c$ this term can be associated with an $\ve_Z^a$ variation of $e^a$. Therefore, on the mass shell, the local Poincar\'e translations are a redundant symmetry and can be completely substituted with the $3d$ diffeomorphisms, while off the mass shell the local Poincar\'e translations can be used to set $\phi^a=0$. Note that once this is done the flat space one-forms \eqref{CF} are obtained from \eqref{One-forms} by simply setting $e^a=dx^a$ and ${h}^a=0$.

We are now ready to generalize the actions \eqref{S1}-\eqref{3L} to describe gauge-invariant couplings of the Goldstone field $A_a(x)$ to the spin-2 fields $e^a$, ${h}^a$ and $\omega^a$ by replacing $E^a_0$ and ${H}_0^a$ with $E^a$ and ${H}^a$ defined in \eqref{One-forms}. We thus get the following symmetry breaking action 
\be\label{Action_Matter}
S_{sym. br.}= \frac 12\int_{\mathcal{M}_3} \ve_{abc}\left(\frac{\Lambda_0}3 E^a E^b E^c+\tilde\b E^a E^b {H}^c+\tilde\alpha E^a {H}^b {H}^c+\frac{\tilde\rho}3 {H}^a {H}^b {H}^c\right),
\ee
where $\Lambda_0$, $\tilde\b$, $\tilde\alpha$ and $\tilde\rho$ are arbitrary  coupling constants whose dimensions are determined by appropriate powers of the symmetry breaking parameter $f$.
Note that the first Volkov-Akulov-like term in \eqref{Action_Matter} generates a cosmological constant. Note also that, in contrast to \eqref{3L}, the last two terms in \eqref{Action_Matter} are not total derivatives.

We will now show that the theory described by the sum of the actions (\ref{Action_Gravity}) and \eqref{Action_Matter} contains the Minimal Massive Gravity of \cite{Bergshoeff:2014pca}. The MMG and HMCSG actions are related to each other by a linear transformation of the three spin-2 fields when certain parameters in the latter are set to zero.

\section{From spontaneously broken HMCSG to MMG}
The action \eqref{Action_Matter} contains the Goldstone fields $\phi^a$ and $A^a$ which, as usual, can be gauge fixed to zero by the corresponding local symmetry transformations \eqref{dfA} with the  parameters $\ve^a_P=-\phi^a$ and $\ve^a_Z=-A^a$. In this (unitary) gauge the one-forms $E^A$ and ${H}^a$ reduce, respectively, to $e^a$ and ${h}^a$, and we get the gauge-fixed action
\bea\label{Action_Gravity+M}
S_{HMCSG}&=& \frac 12\int_{\mathcal{M}_3} \,\left(-\sigma\,(2e^a R_a+{h}^{a}\nabla {h}_{a})+\frac 1 { m}(\omega^a d\omega_a+\frac 13\varepsilon_{abc}\omega^{a}\omega^b\omega^c)\right)\nonumber\\
&+&
 \frac 12\int_{\mathcal{M}_3} \ve_{abc}\left(\frac{\Lambda_0}3\, e^a e^b e^c+\tilde\b\, e^a e^b {h}^c+\tilde\alpha e^a {h}^b {h}^c+\frac{\tilde\rho}3\, {h}^a {h}^b {h}^c\right) ,
\eea
whose residual symmetries are the $3d$ local Lorentz transformations and the diffeomorphisms.

On the other hand, in our conventions and notation the MMG action \cite{Bergshoeff:2014pca} has the following form
\bea\label{MMG}
S_{MMG}&=& \frac 12\int_{\mathcal{M}_3} \,\left(-2\sigma e^a R_a+2\,{h}^{a}\nabla e_{a}+\frac 1 {m}(\omega^a d\omega_a+\frac 13\varepsilon_{abc}\omega^{a}\omega^b\omega^c)\right)\nonumber\\
&+&
 \frac 12\int_{\mathcal{M}_3} \ve_{abc}\left(\frac{{\Lambda_0}}3\, e^a e^b e^c+ \alpha\, e^a {h}^b {h}^c\right),
\eea
where again $\sigma=\pm 1$, and $m$, $\Lambda_0$ and $\alpha$ are arbitrary (dimensionful) parameters, and the spin-2 fields are formally denoted in the same way as in \eqref{Action_Gravity+M} to simplify notation, though now ${h}^a$ is dimensionless. Note that when $\alpha=0$ in \eqref{MMG}, the action reduces to the first-order action for the Topologically Massive Gravity (TMG) \cite{Deser:1981wh} for which the requirement of positive energy of the massive spin-2 mode singles out the sign $\sigma=-1$, while for  General Relativity $\sigma=1$ (see the discussion in \cite{Bergshoeff:2014pca}).

The difference between the actions \eqref{Action_Gravity+M} and \eqref{MMG} is obvious. However, we will now show that the MMG action is a particular case of \eqref{Action_Gravity+M} with three independent parameters.  To this end, it is useful to notice that the actions are of a Chern-Simons-like type \cite{Hohm:2012vh,Bergshoeff:2014pca,Merbis:2014vja}, i.e. they can be written in the following form
\be\label{Action CS-like}
S = \int_{\mathcal{M}_3} \left(\frac12 g_{rs}a^r\cdot d a^s+\frac16 f_{rst} a^r\cdot a^s\times a^t\right),
\ee
where $a^{ra}=(e^a,{h}^a, \omega^a)$ (i.e. $r=1,2,3$ stand, respectively, for $r=e,{h},\omega)$, and $g_{rs}$ and $f_{rst}$ are symmetric tensors with constant components.  In \eqref{Action CS-like}  we used the convenient $3d$ Lorentz-vector algebra notation \cite{Hohm:2012vh}
\be
(a^r\times a^s)^a=\ve^{abc} a^r_b a^s_c, \qquad a^r\cdot a^s=\eta^{ab}a^r_a a^s_b \, .
\ee

In the case of \eqref{Action_Gravity+M}  $g_{rs}$ and $f_{rst}$ have the following non-zero components
\bea\label{gf}
&&
g_{e\omega}=- \sigma, \qquad  g_{\omega\omega}=\frac{1}{m}, \qquad g_{{h}{h}}=-\sigma,\nonumber\\[2pt]
&&
f_{eee}=\Lambda_0, \qquad f_{\omega\omega\omega}=\frac{1}{m}, \qquad f_{{h}{h}{h}}=\tilde\rho,
\nonumber\\[2pt]
&&
f_{e e {h}}=\tilde\b, \qquad f_{e \omega \omega}=- \sigma, \qquad f_{e{h}{h}}=\tilde\alpha, \qquad f_{\omega{h}{h}}=-\sigma,
\eea
while for MMG \eqref{MMG}
\bea\label{MMGgf}
&&
 g_{e\omega}=- \sigma, \qquad  g_{\omega\omega}=\frac{1}{m}, \qquad g_{e{h}}=2,\nonumber\\[2pt]
&&
{f}_{e \omega \omega}=- \sigma,  \qquad {f}_{\omega\omega\omega}=\frac{1}{m}, \qquad { f}_{{e}{h}\omega}=1\,,
\nonumber\\[2pt]
&&
{f}_{eee}={\Lambda_0}, \qquad { f}_{e{h}{h}}=\alpha.
\eea
The matrix of the linear transformation of the fields
\be\label{aTa}
\tilde a^p=T^p{}_q a^q,
\ee
which relates (modulo a total derivative) the HMCSG tensor $g_{pr}$ in \eqref{gf} to the MMG one in \eqref{MMGgf}
\be\label{gg}
g_{{}_{MMG}}=T^T g_{{}_{HMCSG}}\,T,
\ee
has the following form
\be\label{T-1}
T^p{}_q= \begin{pmatrix}
    1 & -\frac{1}{m} & 0\\
    0 & \frac 1{\sqrt{-m\sigma}} & 0\\
    0 & -\sigma & 1
\end{pmatrix}.
\ee
Note that the form of the matrix $T$ requires $m\sigma$ to be negative. This is related to the sign of $g_{{h}{h}}=-2\sigma$ in the HMCSG case. This sign can be flipped by performing the parity transformation $e^a \to -e^a$ and $\sigma \to -\sigma$ in the HMCSG action \eqref{Action_Gravity+M}.

Thus, upon performing the transformation \eqref{aTa}
one brings the action \eqref{Action_Gravity+M} to the following form (in which, for simplicity, we remove `tilde' over the redefined fields)
\bea\label{HMCSGa}
S_{HMCSG}&=&\frac 12\int_{\mathcal{M}_3} \,\left(-2\sigma e^a R_a+2\,{h}^{a}\nabla e_{a}+\frac 1 {m}(\omega^a d\omega_a+\frac 13\varepsilon_{abc}\omega^{a}\omega^b\omega^c)\right)\nonumber\\
&+&
\frac 12\int_{\mathcal{M}_3} \ve_{abc}\left(\frac{{\Lambda_0}}3\, e^a e^b e^c+ \alpha\, e^a {h}^b {h}^c+\beta\, e^a e^b {h}^c+ \frac{\rho}3\, {h}^a {h}^b {h}^c \right),
\eea
where
\bea\label{tilde}
 \beta=\frac{\tilde\beta}{\sqrt{-m\sigma}}-\frac{{\Lambda_0}}m,\qquad \alpha=-\frac{2\tilde \beta}{m\sqrt{-m\sigma}}-\frac{\tilde \alpha\sigma}m+\frac{{\Lambda_0}}{m^2}{-\s},\nonumber\\ [2pt]
\rho=-\frac{\tilde\rho\sigma}{m\sqrt{-m\sigma}}+\frac{3\tilde\beta}{m^2\sqrt{-m\sigma}}+\frac{3\tilde\alpha\sigma}{m^2}-\frac{{\Lambda_0}+m^2\sigma}{m^3}\,
\eea
and the values of $g_{rs}$ and $f_{rqs}$ are
\bea\label{MMGgftilde}
&&
 g_{e\omega}=-  \sigma, \qquad  g_{\omega\omega}=\frac{1}{m}, \qquad g_{e{h}}= 1,\nonumber\\[2pt]
&&
{f}_{e \omega \omega}=- \sigma, \qquad {f}_{\omega\omega\omega}=\frac{1}{m}, \qquad { f}_{{e}{h}\omega}=1 \,,\qquad
\nonumber\\[2pt]
&&
{f}_{eee}={\Lambda_0},  \qquad { f}_{e{h}{h}}=\alpha,  \qquad { f}_{e{e}{h}}=\beta\,,\qquad { f}_{h{h}{h}}=\rho\,.
\eea
The action \eqref{HMCSGa} reduces to the MMG action \eqref{MMG}
when $\beta=\rho=0$.

The equations of motion which follow from \eqref{HMCSGa} are
\bea\label{eomnew}
&&
-2\sigma R+2\nabla{h}+{\Lambda_0}\, e\times e+ \alpha{h}\times{h}+{2 \beta\, e\times {h}}=0,
\nonumber\\[2pt]
&&
2\nabla e+{2 \alpha} e \times {h}+ \beta e\times e+{ \rho} {h} \times {h}=0,
\\[2pt]
&&
-2\sigma \nabla e+\frac{2}{ m}R+2 e \times{h}=0.\nonumber
\eea
Note that in order to have three independent dynamical equations, the coefficient of the gravitational Chern-Simons term, i.e. $1/m$,
should be non-zero.
A linear combination thereof brings the above equations to the form
\bea\label{eomnew1}
&&
2R+2 m (1+\sigma \alpha) e \times{h}+{ \sigma m \beta e\times e+{\sigma m \rho}\, {h} \times {h}}=0, \nonumber\\[2pt]
&&
2\nabla{h}
+2 (m\sigma(1+\sigma \alpha)+ \beta)\, e \times{h}+{ (m \beta+{\Lambda_0}) e\times e+({m \rho}+ \alpha) {h} \times {h}}=0,
\nonumber\\[2pt]
&&
2\nabla e+{2 \alpha} e \times {h}+{  \beta e\times e+{ \rho} {h} \times {h}}=0.
\eea
Upon the redefinition of the connection
\be\label{FromOtoo}
\Omega=\omega+{ \alpha} {h}+\frac{ \beta}2\,e,
\ee
we have
\bea\label{eomnew2}
&&
2 R(\Omega)+C_1 e\times e+C_2\, e\times{h}+\frac \rho 2 C_3\, h\times h =0,\nonumber
\\[2pt]
&&
2\nabla(\Omega){h}+C_3 e\times{h}+ (\Lambda_0+m\beta)e\times e +(m\rho - \alpha){h}\times{h}=0,\nonumber\\[2pt]
&&
2\nabla(\Omega)e+\rho h\times h=0,
\eea
where
\bea\label{Ci}
C_1&=&\frac 14\left(\beta +2m\sigma\right)^2+\alpha(\Lambda_0+m\beta)-m^2,
\nonumber\\[2pt]
C_2&=&2\left(\alpha\left(\beta+ 2m\sigma\right)+m(1+\alpha^2)\right)
\\[2pt]
 C_3&=& (\beta+ 2m\sigma)+2m\alpha.\nonumber
\eea
Note that in \eqref{eomnew2} (and \eqref{Ci}) $\beta$ always appears in the combinations $\Lambda_0+m\beta$ and $ \beta +2m\sigma$. So effectively $\beta$  shifts $\Lambda_0$ and promotes $\sigma=\pm 1$ to a fully-fledged continuous parameter that cannot be scaled away and may take zero value.

Taking the covariant derivative of these equations and comparing the results one finds that for consistency either
\be
 \alpha\left(\beta+ 2m\sigma\right)+m(1+\alpha^2) - \r({\Lambda_0}+m \b)=0,
\label{choice1}
\ee
or
\be\label{Constraint_From equations}
{h}\cdot e=0.
\ee
The latter implies that ${h}^a_\m e^a_\n$ is a symmetric tensor as in the MMG theory \cite{Bergshoeff:2014pca}, for which the first option \eqref{choice1} reduces to
\be\label{choice1MMG}
1+ \alpha \sigma=0\,.
\ee
For $\rho=0$ the equations \eqref{eomnew2} take the form
\bea\label{eomnew22}
&&
2 R(\Omega)+C_1 e\times e+C_2\, e\times{h} =0,\nonumber
\\[2pt]
&&
2\nabla(\Omega){h}+C_3 e\times{h}+ (\Lambda_0+m\beta)e\times e  - \alpha{h}\times{h}=0,\nonumber\\[2pt]
&&
2\nabla(\Omega)e=0,
\eea
Note that now among the five coefficients $C_i$ $(i=1,2,3)$, $\Lambda+m\beta$ and $\alpha$ only four are functionally independent and expressed in terms of four independent parameters $\Lambda_0+m\beta$, $\alpha$, $\beta+2m\sigma$ and $m$.

We see that when $ \rho=0$ the geometry is torsionless and, in addition, the first equation can be solved for ${h}$ as in MMG (provided that
$C_2=\alpha(\b+2m\sigma) +m(1+ \alpha^2) \neq 0$ and hence \eqref{Constraint_From equations} is satisfied), but in our case there is still one more independent coupling constant $ \beta$ like in \cite{Afshar:2019npk} (equations (A5)-(A8) therein).\footnote{Also, when $  \rho$ is non-zero, one can make a shift $e\to e+c{h}$ (with an appropriate constant $c$) such that for a certain range of the parameters  the term ${h}\times {h}$ disappears from the first equation of \eqref{eomnew2}. Thus, one can solve it for ${h}$, but the geometry, in general, remains torsionful, due to the structure of the last two equations in \eqref{eomnew2}. So it is not possible, in general, to solve these equations for $\Omega$ in terms of the dreibein $e$. Still, as we will see below, also in the case with $\rho\not =0$ the theory has a single propagating bulk degree of freedom and can be studied perturbatively around an $AdS_3$ vacuum, like the MMG.} Alternatively, if we would like to treat ${h}$ as the dreibein, we can arrive at the torsionless condition by modifying the connection  starting from the second equation in \eqref{eomnew1} and setting \hbox{${\Lambda_0}=0$}.

As in the MMG case \cite{Bergshoeff:2014pca}, solving the first equation in \eqref{eomnew22} for $h$ we get
\be\label{h=}
h_{\mu\nu}=h^a_{\mu}e^b_\nu\eta_{ab}=-\frac{2}{C_2}\left(S_{\mu\nu}+\frac{C_1}2 g_{\mu\nu} \right), \qquad g_{\mu\nu}=e^a_\mu e^b_\nu\eta_{ab},
\ee
where $S_{\mu\nu}=R_{\mu\nu}-\frac 14 g_{\mu\nu}R$ is the $3d$ Schouten tensor. Substituting this solution into the second equation of \eqref{eomnew22} and expressing $\Omega^a$ through $e^a$ by solving the torsionless condition in \eqref{eomnew22} we get
\be
\label{MetricFieldeq}
C_{\m\n} +  \left( \frac{C_3}{2} + \frac{\alpha C_1}{C_2} \right) G_{\m\n}
- \frac 12\left( {C_3 C_1} - {(\Lambda_0 +m\beta)  C_2} +\frac{\alpha C_1^2}{C_2}\right) g_{\m\n} = \frac{2\alpha }{C_2} J_{\m\n} ,
\ee
where $G_{\m\n}$ is the Einstein tensor, $ C_{\m\n}= \frac{1}
{\sqrt{-\textrm{det}\, g}} \varepsilon_{\m}^{\,\, \tau \rho}
\nabla_{\tau} S_{\rho\n}
$ is the Cotton tensor
and $J_{\m\n}= \frac{1}{2\,\textrm{det}\, g} \varepsilon_{\m}^{\,\, \rho \sigma}
\varepsilon_{\n}^{\,\, \tau \eta} S_{\rho\tau}S_{\sigma\eta}$. The above equation has the same form as the MMG metric field equation  \cite{Bergshoeff:2014pca} containing three coefficients, which are now composed of four continuous parameters  $\Lambda_0+m\beta$, $\alpha$, $m$ and $\beta+2m\sigma$.

\section{$SL(2,R) \times SL(2,R)\times SL(2,R)$ CS theory as a degenerate case of MMG and HMCSG}\label{SL3}
Though the main subject of this  paper is the massive gravity theory whose fields satisfy the consistency condition \eqref{Constraint_From equations}, in this Section we would like to elucidate the structure of the model for which the equation \eqref{choice1} holds, so the model has only five independent parameters. Then the equations \eqref{eomnew1} (or \eqref{eomnew2}) are integrable in the sense that their covariant derivatives are identically zero without imposing the additional constraint \eqref{Constraint_From equations} on the fields. This means that \eqref{eomnew1} become the Maurer-Cartan equations for the  one-forms $e^a$, ${h}^a$ and $\omega^a$ which should thus be the components of a Cartan form associated with a gauge group of rank 9. This group is semi-simple and should contain the $3d$ Lorentz group $SL(2,R)$ as a subgroup. As such, the most reasonable candidate is $SL(2,R) \times SL(2,R)\times SL(2,R)$. A Chern-Simons gravity based on this group was considered in \cite{Diaz:2012zza,Fierro:2014lka,Concha:2018jjj}.

To show that this is indeed so, let us  consider a simpler case in which $ {\rho}=0$. Then in eq. \eqref{eomnew1}, in which the remaining parameters satisfy the condition \eqref{choice1}, we redefine the fields $e^a$ and $\omega^a$ as follows
\bea\label{to1}
&&
{h}\to\frac 1{ \alpha}{h}+\frac{m}{ \alpha}\left( \alpha^2-1\right)\left(4{\Lambda_0} \alpha^3-m^2(1+ \alpha\s)^3(3 \alpha\s-1)\right)^{-1/2} e,
\nonumber\\[2pt]
&&
\omega\to \omega+2m(1+ \alpha\s)\left(4{\Lambda_0} \alpha^3-m^2(1+ \alpha\s)^3(3 \alpha\s-1)\right)^{-1/2}e- {h},\\[2pt]
&&
e\to {2 \alpha }(4{\Lambda_0} \alpha^3-m^2(1+ \alpha\s)^3(3 \alpha\s-1))^{-1/2} e\,,\nonumber
\eea
where we assume, without loss of generality, that the expression under the square root is positive.
Then the equations \eqref{eomnew1} (with $ \rho=0$ and $  \beta=-\frac m{ \alpha}(1+\sigma\, \alpha)^2$) take the following form
\bea\label{MMG2}
&&
R+\frac12 e\times e=0,
\nonumber\\[2pt]
&&
\nabla e=0,
\\[2pt]
&&
\nabla{h}-\frac12 {h}\times{h}+\frac12 e\times e=0.\nonumber
\eea
As one can easily check, these are the Maurer-Cartan equations for the one-form \linebreak ${\mathbf A}=\omega^a J_a +e^a P_a +{h}^a Z_a$ associated with the following linear combinations of the three sets $T_1$, $T_2$ and $T_3$ of the generators of $SL_1(2,R) \times SL_2(2,R) \times SL_3(2,R)$, respectively:
\be\label{sl2r3}
J=T_1+T_2+T_3, \qquad P=T_1-T_2, \qquad Z=-T_3\,.
\ee
In the general case (i.e. when $ \rho\not =0$) the transformation of the fields to the form which results in eq. \eqref{MMG2} is much more cumbersome and we will not give it here.

We have thus found that the action \eqref{HMCSGa} which produce the equations of motion \eqref{eomnew1} with the parameters satisfying the condition \eqref{choice1} is similar to that of \cite{Concha:2018jjj}.
Therefore in this case all the bulk degrees of freedom are pure gauge, as e.g. in the case of Gravity based on $SL(2,R) \times  SL(2,R)$.
Here we just have an additional $SL(2,R)$ field. Of course, the physical content of the theory depends on the boundary conditions which can be imposed on the components of $\mathbf A$. These boundary conditions determine for us which is the true dreibein and connection and asymptotic symmetries. For instance, we can associate them with those belonging to $SL_1(2,R) \times SL_2(2,R)$ and then the third $SL_3(2,R)$ gauge field completely decouples (see \cite{Diaz:2012zza,Fierro:2014lka,Concha:2018jjj} for more details).

In summary, the particular choice of the parameters \eqref{choice1} in the HMCSG action does not break the Hietarinta/Maxwell symmetry but deforms it to $SL(2,R) \times SL(2,R)\times SL(2,R)$. This is similar to how the Poincar\'e symmetry gets deformed to the (A)dS symmetry by adding the cosmological term to the Einstein gravity action. On the other hand, since the Hietarinta/Maxwell algebra is a contraction of the $sl(2,R) \times sl(2,R)\times sl(2,R)$ algebra, the HMCS action \eqref{Action_Gravity0} can be obtained as the contraction limit of the $SL(2,R) \times SL(2,R)\times SL(2,R)$ Chern-Simons action.

%%%%%%%%%%%%%%%%%%%%%%%%%%%%

\section{Hamiltonian analysis}\label{Ha}

We shall now  sketch, following \cite{Hohm:2012vh,Bergshoeff:2014pca,Bergshoeff:2014bia}, the Hamiltonian analysis of the system described by the action \eqref{Action_Gravity+M} and show that it has one propagating degree of freedom as in the particular case of the MMG model.

Let us assume that the manifold $\mathcal{M}_3$ on which the theory is defined can be presented as the product $\mathbb{R}\times \Sigma$, where $\Sigma$ is a two-dimensional manifold with boundary parametrized by the coordinates $x^i$, $i=1,2$, while $\mathbb{R}$ defines the temporal direction parametrized by $x^0$.
Upon this splitting the general Chern-Simons-like action \eqref{Action CS-like} takes the following form
\be\label{splitform}
S = \int_{\mathbb{R}}dx_0 \int_{\Sigma} d^2x \varepsilon^{ij}\left[g_{rs}\dot{a}_{i}^r\cdot a_{j}^s+a_{0}^r\cdot \Big(g_{rs}\partial_i a^{s}_j+\frac12 f_{rst} a^{s}_{i}\times a^{t}_{j}\Big)\right],
\ee
where dot denotes the derivative with respect to $x^0$ and $\varepsilon^{0ij}\equiv\varepsilon^{ij}$.

From the form of this action we see that the canonical momenta $p^i_{ar}$ associated to $a_i^{ar}$ are constrained to be linear combinations of the fields themselves
$$
p^i_{ar}=\varepsilon^{ij}g_{rs} a_{j}^{as}\,.
$$
Upon solving these constraints, one gets the equal-time Poisson (actually Dirac) brackets for the fields $a^{ar}_i$
$$
\{a^{ar}_i(x),a^{bq}_j(y)\}=\varepsilon_{ij}\eta^{ab}g^{rq}\delta^2(x-y)\,,
$$
where $g^{rq}$ is the inverse of $g_{rq}$.

From the structure of \eqref{splitform} we also see that $a_{0}^s$ plays the role of a Lagrange multiplier giving rise to 9 constraints
$$\varphi_r^a=\varepsilon^{ij}
\Big(g_{rs}\partial_i a^{s}_j+\frac12 f_{rst} a^{s}_{i}\times a^{t}_{j}\Big)^a. $$
The corresponding constraint functional for arbitrary fields $\chi^r_a$ with well defined variation has the following form
\be\label{Constraints}
\varphi[\chi]=\int_{\Sigma}d^2x \chi^r\cdot  \varepsilon^{ij}\Big(g_{rs}\partial_i a^{s}_j+\frac12 f_{rst} a^{s}_{i}\times a^{t}_{j}\Big)+\int_{\partial\Sigma}dx^i \chi^r\cdot a_r.
\ee
The Poisson brackets of these constraints have the following structure
\be\label{Constraints_Bracket}
\{\varphi[\chi],\varphi[\xi]\}=\varphi[[\chi,\xi]]+\int_{\Sigma}d^2x \chi^r_a\xi^s_b \mathcal{P}^{ab}_{rs}-\int_{\partial \Sigma}d\phi \chi^{r}\cdot (g_{rs}\partial_\phi\xi^{s}+f_{rst}a^{s}_\phi\times\xi^t),
\ee
with $[\chi, \xi]_t=f_{rst}\chi^r\times\xi^s$ and
\be\label{Matrix_P}
\mathcal{P}^{ab}_{rs}=f^t_{q[r}f_{s]pt}\eta^{ab}\Delta^{pq}+2f^t_{r[s}f_{q]pt} V^{ab, pq}, \qquad V_{ab}^{pq}=\varepsilon^{ij}a^p_{ia}a^q_{jb}, \qquad \Delta^{pq}=V_{ab}^{pq}\eta^{ab}.
\ee
The  integration variable $\phi$ parametrises a (compact) boundary $\partial\Sigma$.

The number of first- and second-class constraints for the model under consideration can be read off from the rank of the matrix $\mathcal{P}$, eq. \eqref{Matrix_P}, in which one should insert the explicit expressions \eqref{MMGgftilde} for the tensors $g_{rs}$ and $f_{rqs}$. Note that in \eqref{Matrix_P} the indices are raised with the matrix $g^{rs}$
\if{}
$$
\begin{pmatrix}
   0  & \frac 12 &0\\
    \frac 12 & \frac m2  & \frac{m\sigma}2\\
    0 & \frac{m\sigma}2 & \frac{m}2
\end{pmatrix}.
$$
\fi
If we assume that (\ref{Constraint_From equations}) holds, we have an additional (secondary) constraint
\be\label{Constaint_Second}
\Delta^{e{h}}=0.
\ee
Taking this into account, a straightforward computation shows that the first term in (\ref{Matrix_P}) vanishes and $\mathcal P$ becomes degenerate
\be\label{Matrix_Constraints bracket}
\mathcal{P}= \left( \alpha \b- \r({\Lambda_0}+m \b)+m(1+ \alpha\s)^2\right) \begin{pmatrix}
   -V_{ab}^{hh}  & V_{ab}^{{h}e}&0\\
    V_{ab}^{e{h}} & -V_{ab}^{ee}  & 0\\
    0 & 0 & 0
\end{pmatrix}.
\ee

Now one should also compute the Poisson brackets of the constraint \eqref{Constaint_Second} with  $\varphi(\chi)$. Using a general formula of \cite{Hohm:2012vh} one gets
\bea
&&
\{\Delta^{e{h}}, \varphi[\chi]\}=\varepsilon^{ij}\left(\nabla_i\chi^{e}\cdot{h}_j-\nabla_i\chi^{h}\cdot{e}_j\right)+\varepsilon^{ij}e_i \times{h}_j\cdot\left(m \rho\chi^{{h}}+m\s(1+ \a\s) \chi^{e}\right)
\nonumber\\[2pt]
&&
-\varepsilon^{ij}{h}_i\times{h}_j\cdot\left( \alpha\chi^{e}+ \rho\chi^{{h}}\right)
+\varepsilon^{ij}e_i\times e_j\cdot\left((\Lambda_0+m \b+m\s(1+ \a\s))\chi^{e}+ \b \chi^{{h} }\right)
\eea
with
\be
\nabla_i\chi=\partial_i\chi+\omega_i\times\chi.
\ee

As in the MMG  \cite{Bergshoeff:2014pca} we thus have the $(10\times10)$ matrix of the Poisson brackets of 10 constraints, i.e. $\varphi^{a}_r$ and $\Delta^{e{h}}$, which has rank four. This implies that, if the coefficient in front of the matrix \eqref{Matrix_Constraints bracket} is non-zero, system has 6 first-class and 4 second class constraints which reduce the number of the phase-space physical degrees of freedom in $a^{ar}_i$ to 2, i.e the system has a single bulk degree of freedom in the Lagrangian formulation.

When the coefficient in \eqref{Matrix_Constraints bracket} is zero,
which is equivalent to the choice \eqref{choice1},
the constraint \eqref{Constaint_Second} is absent and one has 9 first-class  constraints $\varphi^{a}_r$ which reduce the number of bulk physical degrees of freedom to zero. In this case, as we discussed in Section \ref{SL3}, the considered system reduces to the
Chern-Simons theory with the gauge group $SL(2,R)\times SL(2,R)\times SL(2,R)$.

\section{$AdS_3$ background and the central charges of the asymptotic symmetry algebra}

We shall now study properties of the HMCSG theory for field configurations whose geometry is asymptotically $AdS_3$ and compute the corresponding centrally extended asymptotic symmetry algebra which underlies a dual $CFT_2$.

\subsection{$AdS_3$ solution of the HMCSG field equations}
For the $AdS_3$ background to satisfy the field equations \eqref{eomnew2} we take the following ansatz for the vevs of $e$, ${h}$ and $\Omega$
\be\label{AdS3}
\langle e\rangle:=\bar e\,, \qquad \langle {h} \rangle:= mC\bar e\,,\qquad \langle \Omega \rangle:=\bar\Omega- \frac{ \rho m^2 C^2}2\bar e\,,
\ee
where $\bar e$ and $\bar\Omega$ are $AdS_3$ dreibein and connection, and $C$ is a real dimensionless parameter.

Substituting this ansatz into equations \eqref{eomnew2} we find that, provided that $C$ satisfies the cubic equation
\be\label{C=}
 \rho m^3 C^3-( m \r  - \alpha)m^2C^2-( \b+2m\s(1+ \alpha\s))\,mC-({\Lambda_0}+ m \b  )=0,
\ee
which always has at least one real root,
eqs. \eqref{eomnew2} reduce to those describing the $AdS_3$ space
\be\label{Equations_vacuum}
\bar R(\bar\Omega)+\frac{l^{-2}               } 2\bar e\times \bar e=0, \qquad \bar\nabla \bar e=0,
\ee
where
\bea\label{Lambda}
l^{-2} \equiv -\Lambda&=&\frac{{ \rho}^2m^4C^4}4 +\frac{ \r m^2 C^2}{2}( \b+ 2 m \s(1+ \alpha\s))+2mC( \b \alpha+m(1+ \alpha\s)^2)\nonumber\\
&&+\left(\frac{{  \b}^2}{4}+{\Lambda_0} \alpha+ m \b\s(1+ \alpha\s)\right)\,\nonumber\\
 &=&\frac 14\left({{ \rho}m^2C^2}+ \b+ 2 m \s(1+ \alpha\s)\right)^2+2mC( \b \alpha+m(1+ \alpha\s)^2)\nonumber\\
&&+{\Lambda_0} \alpha- m^2(1+ \alpha\s)^2\,.
\eea
$l^{-2}$ is assumed to be positive so that the cosmological constant $\Lambda$ is negative.\footnote{A more general class of vacuum solutions in MMG including those with a positive cosmological constant were considered e.g in  \cite{Arvanitakis:2014yja, Alishahiha:2014dma, Giribet:2014wla, Arvanitakis:2015yya, Altas:2015dfa, Deger:2015wpa, Charyyev:2017uuu, Sarioglu:2019uxv}, in particular at a specific point called  ``merger point". The merger point is a point in the space of the parameters of the theory at which for all values of $C$ defined by the equation  \eqref{C=} the cosmological constant $\Lambda$ \eqref{Lambda} has a unique value. It would be of interest to study a similar class of vacuum solutions also in the HMCSG context.}

\subsection{Asymptotic symmetries and central charges}
In \cite{Brown:1986nw} Brown and Henneaux studied asymptotic symmetry properties of the pure $3d$ GR with $AdS_3$ boundary conditions.
The local $3d$ Lorentz symmetry and $3d$ diffeomorphisms of GR give rise to  six first-class constraints generating these symmetries. These can be split into two mutually commuting sets of generators corresponding to the $SL(2,R) \times SL(2,R)$ group of the Chern-Simons formulation of the theory. When evaluated on an asymptotically $AdS_3$ space, each set was shown to generate the Virasoro algebra with a nontrivial central extension. This analysis was generalized to 3D massive theories of gravity in \cite{Carlip:2008qh, Bergshoeff:2014pca, Compere:2009zj, Bergshoeff:2013xma, Merbis:2014vja} and to the Maxwell-Chern-Simons gravity in \cite{Concha:2018zeb}.

We will now carry out the computation of the centrally extended asymptotic symmetry algebra for the HMCSG theory, following closely the steps explained in detail in \cite{Carlip:2008qh} and \cite{Bergshoeff:2014pca}. Consider the following combination of the constraints \eqref{Constraints}
\be\label{Virasoro_generaors}
L_{\pm}[\chi]=\varphi_e[\chi^\m e_\m]+\varphi_h[\chi^\m h_\m]+a_{\pm}\varphi_\omega[\chi^\m e_\m],
\ee
in which the parameters in the brackets are field-dependent and  $\chi^\mu(x)$ are associated with the parameter of $3d$ diffeomorphisms.

For convenience we have defined $\varphi_\omega$ for the spin connection $\omega$ in  \eqref{HMCSGa}.
The constant parameters $a_\pm$ should be properly tuned in order to make the Poisson bracket of $L_+$ and $L_-$ vanish. It can be shown that \eqref{Virasoro_generaors} are a combination of the first-class constraints, corresponding to the local $3d$ Lorentz transformations and diffeomorphisms \cite{Carlip:2008qh,Bergshoeff:2014pca}. Using the general formula \eqref{Constraints_Bracket} one finds that for the $AdS_3$ solution \eqref{AdS3} the Poisson bracket of $L_+$ and $L_-$ reduces to
\bea\label{PB}
&&
\{L_+[\chi], L_-[\eta]\}=\varphi_\omega[[\chi,\eta]]\left(a_+ a_-+2m^2C(1+ \a\s)+m \b\s+m^3 C^2 \r\s\right)
\nonumber\\[2pt]
&&
\qquad \qquad +\left(\varphi_e[[\chi,\eta]]+mC\varphi_h[[\chi,\eta]]\right) \left(a_++a_-+2 \a m C+ \b+m^2 C^2  \r\right).
\eea
Note that on the $AdS_3$ solution \eqref{AdS3} the second term in \eqref{Constraints_Bracket} vanishes. Also the boundary contribution (the last term in \eqref{Constraints_Bracket}) vanishes. To see this, one should take into account the linear redefinition which relates $\Omega$ with $\omega$  \eqref{FromOtoo}, the vacuum value of the $\Omega$ spin connection \eqref{AdS3}, and the corresponding $AdS_3$ asymptotic symmetry parameters $\chi$ and $\eta$ (see \cite{Carlip:2008qh} for details).
Requiring that the Poisson bracket \eqref{PB} vanishes, we find that the parameters $a_{\pm}$ should have the following values
\be
a_\pm=\pm\frac{1}{l}- \a m C-\frac{ \b}{2}-\frac{C^2 m^2 \r}{2},
\ee
where $l$ is the radius of the $AdS_3$ background defined in \eqref{Lambda}. Using the general expression \eqref{Constraints_Bracket} once again, one also finds \bea\label{BCFT}
&&
\{L_\pm[\chi], L_\pm[\eta]\}=\pm\frac{2}{l}L_\pm[[\chi,\eta]]
\\[2pt]
&&
\pm\frac{2}{l}\left(\s\pm\frac{1}{m l}+ \a C+\frac{ \b}{2m}+\frac{m \r C^2}{2}\right)\int_{\partial\Sigma}d\phi\chi\cdot\left(\partial_\phi\eta+\bar\Omega_\phi\times\eta\pm\frac{1}{l}\bar e_\phi\times\eta\right), \nonumber
\eea
where in order to get the boundary term expressed via the $AdS_3$ spin connection $\bar\Omega$, we made use of \eqref{FromOtoo} and \eqref{AdS3}. After expanding the asymptotic symmetry parameters $\eta$ and $\chi$ in Fourier modes, the commutation relations above represent two copies of the Virasoro algebra with central charges
\be \label{central}
c_\pm=\frac{3l}{2 G}\left(\pm\frac{1}{m l}+\s+\frac{ \b}{2m}+ \a C+\frac{m \r C^2}{2}\right),
\ee
where to be in agreement with the Brown--Henneaux central charge expression \cite{Brown:1986nw} we have included the Newton's constant by restoring $1/16\pi G$ in the action \eqref{HMCSGa}.

For the boundary CFT associated with \eqref{BCFT} to be unitary both central charges should be positive, which implies
\be \label{centralpositivity}
\s+\frac{ \b}{2m}+ \a C+\frac{m \r C^2}{2}-\frac{1}{|m l|} >0 \, .
\ee
For certain choices of the parameters $ \alpha$, $  \beta$, $ \rho$ and $\sigma=\pm 1$, the above expressions reduce to those of pure GR \cite{Brown:1986nw}, TMG \cite{Carlip:2008qh} and MMG \cite{Bergshoeff:2014pca}.

\section{Linearized theory around an $AdS_3$ background}\label{Lp}
We shall now study, following \cite{Bergshoeff:2014pca,Arvanitakis:2014xna}, the conditions on the parameters of our model for which the propagating mode is neither a tachyon nor a ghost.
To this end let us consider perturbations around the $AdS_3$ vacuum solution which are convenient to take as follows
\be\label{vacuum_expansion}
e=\bar e+k, \qquad \Omega=\bar{\Omega}- \frac{ \rho m^2C^2}2(\bar e+{k})-mC \rho\,p+v, \qquad {h}=mC(\bar e+{k})+p,
\ee
where $k$, $v$ and $p$ denote infinitesimal excitations of the fields. Then,
using the relation \eqref{C=} and the definition \eqref{Lambda} of $l^{-2}$ we get
the linearized equations for \eqref{eomnew2} as
\bea\label{eomLC2}
%&&
%\bar\nabla v+l^{-2} \bar e\times k\nonumber\\
%&&
%+\bar e\times p \left({\color{red}3 C  \r( \b+2m\s(1+ \alpha \s))}+ \b \alpha+m(1+ \alpha\s)^2{\color{red}-}3C^2 \rho( \alpha-3m \rho) {\color{red}-9 C^3  {\rho}^2}
%\right)
% =0,
%\nonumber\\
&&
\bar\nabla v+l^{-2} \bar e\times k +
\bar e\times p \left( \b \alpha+m(1+ \alpha\s)^2
- {\rho}({\Lambda_0}+ m \b)\right)
 =0,
\nonumber\\
&& \bar\nabla p+M\,\bar e\times p=0,
\nonumber\\
&&
\bar\nabla k+\bar e\times v=0,
\eea
where \footnote{Note that in the MMG case (i.e. when $\beta=\rho=0$) the  value $M=0$ defines the merger point \cite{Arvanitakis:2014yja} for the values of the cosmological constant. This, however, is not the case anymore for $\rho\not=0$.}
\be\label{Mg}
M=\frac 12\left( \b+2m\s(1+ \alpha\s)+2mC( m  \r- \alpha)-3m^2C^2 \r\right)\,.
\ee

The integrability condition \eqref{Constraint_From equations} for the above equations reduces to
$$
\bar e\cdot p=0.
$$
Making the redefinition (assuming that
$|\ell M| \neq 1$)
\be\label{redefinition_linear_g}
f_{\pm}=\pm {l^{-1}}k+\frac{ \b \alpha+m(1+ \alpha\s)^2- \rho({\Lambda_0}+m \beta)}{\left({\pm } l^{-1}-M\right)}\,p+v,
\ee
one diagonalizes two of the equations \eqref{eomLC2} and gets
\bea\label{linearg}
& \bar\nabla f_{\pm}\pm l^{-1}\bar e\times f_{\pm}=0, & \nonumber\\
[2pt]
&\bar\nabla p+M\bar e\times p=0.
\eea
The first two equations in \eqref{linearg}  describe the linearized $3d$ Einstein gravity with a cosmological constant and the third equation describes the propagation of the spin-2 mode $p$ with the mass $\mathcal M$ given by
$$
{\mathcal M}^2=M^2-l^{-2}.
$$
In accordance with the general Hamiltonian analysis we thus see that the HMCSG model has exactly the same field content as  the MMG. The no-tachyon condition is \cite{Bergshoeff:2014pca}
\be\label{not}
 M^2-l^{-2}>0.
\ee
Let us now find the form of the action \eqref{HMCSGa} up to the second order in perturbations. Upon taking into account the form of the transformation \eqref{FromOtoo}, the excitations \eqref{vacuum_expansion} and the linear redefinition  \eqref{redefinition_linear_g} one gets
\bea\label{quadaction}
&&
S_{2} = \int_{\mathcal{M}_3} \lambda_+ \left(f_+\bar\nabla f_++l^{-1}\bar e\cdot f_+\times f_+\right)+\lambda_-\left(f_-\bar\nabla f_--l^{-1}\bar e\cdot f_-\times f_-\right)
\nonumber\\[2pt]
&&
\qquad \qquad +\int_{\mathcal{M}_3} \frac{1}{m(1-2C)}\left(p\bar\nabla p+M\bar e\cdot p\times p\right),
\eea
where
\be \label{lambdapm}
\lambda_{\pm}=\frac{1}{2m}\mp \frac{l}{4 m}\left(2 m \s+ \b+2m C  \alpha+m^2 C^2  \r\right).
\ee
The first two terms are two linearized $SL(2,R)$ Chern--Simons terms. Comparing \eqref{central} with \eqref{lambdapm} we see that $c_{\pm}= \pm 3 \lambda_{\mp}/G$.

The product of $\lambda_{+}$ and $\lambda_-$ is
\be\label{product}
\lambda_+\lambda_-=-\frac{l^2}{4}  (1-2C).
\ee
If the product is negative, the first two terms describe the linearized pure GR as the difference of two $SL(2,R)$ Chern-Simons terms. In the general case, however, the product may also have the positive sign, then the resulting theory can be interpreted as a kind of ``exotic'' GR with additional  terms.\footnote{Note that the CS action for GR corresponds to the $SO(2,2)$ bilinear form $\langle J^a,P^a\rangle= \langle J_+^a,J_+^a\rangle-\langle J_-^a,J_-^a\rangle=\eta^{ab}$, where $J_\pm^a$ are two copies of $SO(1,2)$ generators, related to that of $SO(2,2)$ as $J^a=J_+^a+J_-^a$ and $P^a=J_+^a-J_-^a$. One can use the additional bilinear form of the $SO(2,2)$ algebra given by $\langle J^a,J^a\rangle=\langle P^a,P^a\rangle=c\,\eta^{ab}$ with a constant $c$ to extend the GR action by the topological and torsion terms $c(\omega d\omega+\frac 13\omega^3)+c e\nabla e$. At the linearized level the sign of the product \eqref{product} depends on the value of the constant parameter $c$.}  However, $-\lambda_+$ and $\lambda_-$ \eqref{lambdapm}
are proportional to the central charges $c_+$ and $c_-$ \eqref{central} in the asymptotic algebra and if we require both central charges to be positive \eqref{centralpositivity}, then the product of $\lambda_{+}$
and $\lambda_{-}$
\eqref{product} must be negative and hence $(1-2C)>0$.
Note that at the chiral point of the
theory, at which one of the boundary central charges vanishes, $1-2C=0$ and equation
\eqref{quadaction} becomes singular.

The last term in \eqref{quadaction}  describes the propagating massive spin-2 mode. The no-ghost condition implies (see \cite{Bergshoeff:2014pca} for details)
\be\label{noghost}
(1-2C) mM<0.
\ee
We shall now consider in more detail three particular cases in which the values of the parameters differ from the original MMG.
\subsection{C=0}\label{C=0}
In this case the equation \eqref{C=} reduces to the following relation
\be
{\Lambda_0}+m \b=0\quad \Rightarrow \quad {\Lambda_0}=- m \beta,
\ee
while \eqref{Lambda} and \eqref{Mg} respectively
simplify to
\be\label{l-2}
l^{-2}=\frac{1}{4}( \b^2+4m \b\s)= m^2\left(\frac{ \b}{2m}+ \s\right)^2-m^2>0 \, ,
\ee
and
\be\label{M}
M= m\left(\frac{ \b}{2m}+\s+ \alpha\right).
%=\frac{2l^{-2}}{ \b}+m( \alpha-\s).
\ee
From \eqref{l-2} we have
\be\label{14}
\frac{ \b}{2m}+\s>1\qquad {\rm or } \qquad\frac{ \b}{2m}+\s<-1\,.
\ee
The no-tachyon condition \eqref{not} takes the form
\be\label{No-tachyon}
%\frac{4m \alpha}{ \b} l^{-2}+m^2(1- \alpha\s)^2>0\, \qquad\Rightarrow \qquad
2 \alpha\left(\frac{ \beta}{2m}+\sigma\right)+1+ \alpha^2>0.
\ee

In the action \eqref{quadaction} we now have
$\lambda_+\lambda_-=-l^{2}/4<0$.
Hence, the first two terms are the difference of two linearized $SL(2,R)$ Chern--Simons terms describing the linearized $3d$ Einstein gravity. The last term describes the propagating massive spin-2 mode whose no-ghost condition \eqref{noghost}
requires
\be\label{mM}
m M<0\qquad \Rightarrow \qquad \frac{ \beta}{2m}+\sigma + \alpha<0.
\ee

The positive central charge condition in the case $C=0$ is
\be\label{pcc0}
\sigma+\frac{ \beta}{2m}-\frac 1{|ml|}>0.
\ee

Now we would like to analyze consequences of the conditions \eqref{14}-\eqref{pcc0}.
From \eqref{pcc0} we see that $\sigma+\frac{ \beta}{2m}>0$ which is compatible with the first choice in \eqref{14}.
Then \eqref{mM} and \eqref{No-tachyon} require that
\be\label{conditions1}
 \alpha < -1, \qquad \left( \alpha+ \frac{ \beta}{2m}+\sigma\right)^2>\left(\frac{ \beta}{2m}+\sigma\right)^2-1\,.
\ee
So finally, the range of the parameters which satisfies the conditions \eqref{14}-\eqref{pcc0} is
\be\label{conditions}
\frac{ \beta}{2m}+\sigma>1,\qquad  \alpha < -\sqrt{\left(\frac{ \beta}{2m}+\sigma\right)^2-1}-\left(\frac{ \beta}{2m}+\sigma\right)\,, \qquad {\Lambda_0}=-{m \beta} \,,
\ee
and $ \rho$ is arbitrary.

%%%%%%%%%%%%%%%%%%%%
\subsection{$ \rho$=0}
In this case we have
\bea\label{rho0}
&& \alpha\,m^2C^2-( \b+2m\s(1+ \alpha\s))mC-({\Lambda_0}+ m \b  )=0, \nonumber\\[2pt]
&&
l^{-2} =2mC( \alpha \b+m(1+ \alpha\s)^2)+\left(\frac{{  \b}^2}{4}+ \alpha{\Lambda_0}+ m \b\s(1+ \alpha\s)\right)\,,
\\[2pt]
&&
M=\frac { \b}2+m\s(1+ \alpha\s)-m \alpha C\,.
\nonumber
\eea
The solution for $C$ is (assuming $  \alpha \neq 0$)
\be\label{Csolution}
C=\frac{ \b+2m\s(1+ \alpha\s)}{2m \alpha}\mp\sqrt{\frac{{\Lambda_0}+ m \b  }{m^2 \alpha}+\frac{( \b+2m\s(1+ \alpha\s))^2}{4m^2 \alpha^2}}\,,
\ee
so
\be\label{Mr0}
M=\pm m \alpha\,\sqrt{\frac{{\Lambda_0}+ m \b  }{m^2 \alpha}+\frac{( \b+2m\s(1+ \alpha\s))^2}{4m^2 \alpha^2}}\,.
\ee
and
\be\label{l2}
l^{-2}=-m^2(1-2C)\left(\frac{ \alpha \b}m+(1+ \alpha\s)^2\right)+M^2>0\,,
\nonumber\\[2pt]
\ee
The no-tachyon condition is
\be\label{nt}
M^2-l^{-2}
=m^2(1-2C)\left(\frac{ \alpha \beta}m+(1+ \alpha\s)^2\right)>0 ,
\ee
and the no-ghost condition is as in \eqref{noghost}. Note that $C$ is real iff $M^2 \geq 0$.  Hence,
the no-tachyon condition also guarantees $C$ to be real. For $M=0$, the $C$ equation \eqref{rho0} has a double root, but this case is un-physical since the no-tachyon condition is violated. Also note that unlike the original MMG, for which $  \beta=0$, the no-tachyon condition does not in general imply $(1-2C)>0$ which is, however, required by the positivity of the asymptotic central charges.

Collecting the positive central charge, the no-tachyon and the no-ghost conditions together we have
\be\label{conditions0}
 \a C+\s+\frac{ \b}{2m}-\frac{1}{|m l|} >0 \quad\Rightarrow \quad 1-2C>0; \qquad \frac{ \alpha \beta}m+(1+ \alpha\s)^2>0, \qquad mM<0 \, .
\ee
The positivity of $l^{-2}$ in \eqref{l2} also requires that
$$
0<1-2C<\frac{M^2}{m^2\left(\frac{ \alpha \b}m+(1+ \alpha\s)^2\right)}.
$$
Due to the definition of $M$ in \eqref{rho0}, the condition $mM<0$ is the same as
\be\label{aC}
 \alpha (C-1)-\frac { \b}{2m}-\sigma>0,
\ee
which when summed up with the first condition in \eqref{conditions0} gives
$$
 \alpha(1-2C)+\frac{1}{|m l|} <0 \qquad \Rightarrow \qquad   \alpha < - \frac{1}{|m l|(1-2C)}<0\,.
$$
From \eqref{aC} and the fact that $ \alpha$ is negative it follows, that in the solution \eqref{Csolution} we should choose the minus sign in front of the square root and plus one in \eqref{Mr0}.

We have thus identified a range of the parameters compatible with the conditions \eqref{conditions0}.
One can proceed further with the analysis and specify this range in more detail. Namely, from the third condition in \eqref{conditions0}, as in the case $C=0$, it also follows that
\be\label{X}
\left(\frac{ \beta}{2m}+\sigma+ \alpha\right)^2 >\left(\frac{ \beta}{2m}+\sigma\right)^2-1.
\ee
\begin{itemize}
\item
Iff
$
|\frac{ \beta}{2m}+\sigma|\geq 1,
$
we have
$$
 \alpha < -\sqrt{\left(\frac{ \beta}{2m}+\sigma\right)^2-1}-\left(\frac{ \beta}{2m}+\sigma\right) \quad \textrm{or} \quad
 \alpha >\sqrt{\left(\frac{ \beta}{2m}+\sigma\right)^2-1}-\left(\frac{ \beta}{2m}+\sigma\right).
$$
These are compatible with $ \alpha<0$ iff $\frac{ \beta}{2m}+\sigma\geq 1$.

\item
Another brunch of \eqref{X} is
$$
-1<\frac{ \beta}{2m}+\sigma<1, \qquad -1< \alpha<0 \, ,
$$
for which a particularly simple case is $\frac{ \beta}{2m}+\sigma=0$.

\item
One more simple case is ${\Lambda_0}=-m  \beta$ for which either $C=0$ and we are back to Subsection \ref{C=0}, or
${ \alpha}C=2\,\left(\frac{ \b}{2m}+\s+ \alpha\right)$ and hence, due to \eqref{aC} and \eqref{nt},
$$
\frac{ \b}{2m}+\s+ \alpha>0, \qquad -1< \alpha<0.
$$
\end{itemize}

Let us now consider the case in which  $  \alpha=0$ (but $ \beta \not=0)$. Then we have
\be\label{gamma=0}
C= - \frac{{\Lambda_0}+ m \b}{m( \b+2m\s)} \, , \quad M=\frac{  \beta}{2} + m \sigma \, , \quad
\ell^{-2} = - \frac{2m ({\Lambda_0}+ m \b  )}{ \b+2m\s} + \frac{  \beta^2}{4} + m\sigma   \beta \, .
\ee
So, the no-tachyon condition is
\be\label{notach}
m^2 + \frac{2m ({\Lambda_0}+ m \b)}{ \b+2m\s} > 0 \, ,
\ee
and the no-ghost condition is
\be\label{nogh}
\left(m+\frac{2 ({\Lambda_0}+ m \b)}{ \b+2m\s}\right)({  \beta} +2 m \sigma) <0 \, .
\ee
From \eqref{notach} and \eqref{nogh} we see that
\be
\frac{  \beta}{2m} +\sigma <0 \, .
\ee
 Note that if $  \beta=0$, the model under consideration
 \eqref{HMCSGa} is TMG \cite{Deser:1981wh} for which the above no-ghost condition requires $\sigma=-1$.

We  will now show that also when $  \beta\not =0$, the model with $ \alpha= \rho=0$ is equivalent to the TMG\cite{Deser:1981wh}. Indeed, upon making the redefinition of the connection \eqref{FromOtoo} in the action \eqref{HMCSGa} with $ \alpha= \rho=0$, we get
\bea\label{TMG+}
S&=&\frac 12\int_{\mathcal{M}_3} \,\left(-\frac{2m\sigma+ \beta}m e^a R_a+\frac 1 {m}(\Omega^a d\Omega_a+\frac 13\varepsilon_{abc}\Omega^{a}\Omega^b\Omega^c)\right)\\
&+&
\frac 16\int_{\mathcal{M}_3}\left({{\Lambda_0}}-\frac{ \beta^3}{8m}\right)\,\ve^{abc}\, e^a e^b e^c+\frac 12\int_{\mathcal{M}_3}\left(2{h}+\frac{ \beta^2+4m \beta\sigma}{4m}\,e^a\right)\nabla e_a\,.\nonumber
\eea
This coincides with the first-order action for the topological massive gravity upon the redefinition $\tilde{h}^a={h}^a+\frac 12\left(\frac{ \beta^2}{4m}+ \beta\sigma\right)\,e^a$ and appropriate rescalings of $e^a$ and $\tilde{h}^a$.

\subsection{$\tilde\alpha=\tilde\beta=\tilde\rho=0$}
Let us now consider the case in which in the action \eqref{Action_Gravity+M} we have $\tilde\alpha=\tilde\beta=\tilde\rho=0$. This is the situation in which the spontaneous breaking of the Hietarinta/Maxwell symmetry occurs only due to the contribution associated with the classical lower-derivative Volokov-Akulov-like Goldstone (``cosmological") term \eqref{S1}. In this case the parameters \eqref{tilde} in the MMG-like action \eqref{HMCSGa} are
\be\label{alphanot0}
  \beta=-\frac{{\Lambda_0}}m, \qquad   \alpha=\frac{{\Lambda_0}}{m^2} -\sigma, \qquad  \rho=-\frac{{\Lambda_0}+m^2\sigma}{m^3}\,.
\ee
Hence, similar to the case of Subsection \ref{C=0} we have ${\Lambda_0}+m \beta=0$ but with particular expressions for $ \alpha$ and $ \rho$ in terms of ${\Lambda_0}$ and $m$. Then the equation \eqref{C=} for $C$ reduces to
\be\label{1C=2}
C\left[{({\Lambda_0}+m^2\sigma)}C^2-2{\Lambda_0} C+{{\Lambda_0}}\right]=0.
\ee
For the solution $C=0$ of this equation from \eqref{Lambda}
and \eqref{Mg} we get
$$
\ell^{-2}=\frac{{\Lambda_0}^2}{4m^2} - {\Lambda_0}\sigma \quad , \quad  M=\frac{{\Lambda_0}}{2m} \, .
$$
We see that to satisfy the no-tachyon \eqref{not} and the no-ghost
\eqref{noghost} conditions together with the requirement $\ell^{-2}>0$
we need $\sigma=-1$ and ${\Lambda_0}<-4m^2$ which are in agreement with \eqref{conditions}. Note that if ${\Lambda_0}=0$,
then $C=0$ and hence this possibility is ruled out by the last inequality. Actually, in this case the background is flat not $AdS$, while the model reduces to the Chern-Simons theory with the unbroken Hietarinta \eqref{Action_Gravity} or Maxwell symmetry  \eqref{Action_Gravity1}.

When $  \rho = 0$, i.e. ${\Lambda_0}=-m^2\sigma$, then from \eqref{1C=2} we see that either $C=0$ or $C=1/2$. In the both cases the no-tachyon condition \eqref{not} is not satisfied.

Finally, let us consider the case in which $C\not =0$, ${\Lambda_0} \neq 0$ and $ \rho \neq 0$. Then from \eqref{1C=2} we get
\be\label{Csoln}
C=\frac{{\Lambda_0} \pm \sqrt{-{\Lambda_0}\sigma m^2 }}{{\Lambda_0}+m^2\sigma}
\ee
So the existence of the real solutions (associated with $AdS_3$ vacua) requires that
\be\label{gs}
{\Lambda_0}\sigma < 0\,.
\ee
In this case from \eqref{Lambda}
and \eqref{Mg} we find
$$
\ell^{-2}=m^2C^2 \quad , \quad  M=\frac{{\Lambda_0}(C-1)}{m} \, .
$$
Note that $C=1$ is not a solution of \eqref{1C=2}.
The no-tachyon condition \eqref{not} becomes
$$
-({\Lambda_0} \sigma +m^2)C^2>0 \quad  \Rightarrow \quad {\Lambda_0} \sigma  <-m^2 \, ,
$$
and the no-ghost condition \eqref{noghost}
using \eqref{1C=2} implies
$$
({\Lambda_0} + m^2 \sigma)(1-C)<0 \, .
$$
Combining the last two inequalities we see
\be\label{1-C}
\sigma(1-C)>0 \, ,
\ee
which shows that we should take the plus sign in the
solution of $C$ in \eqref{Csoln}.

From \eqref{product} the positivity of the central charges implies $1-2C>0$ and since $-{\Lambda_0}\sigma>0$, the no-ghost condition takes the form
$$
{\Lambda_0} (C-1)<0 \quad \Rightarrow \quad \sigma (-{\Lambda_0}\sigma)(1-C)<0 \quad \Rightarrow \quad \sigma(1-C)<0 \, ,
$$
which is incompatible with \eqref{1-C}. Therefore, for the choice of the parameters considered in this Subsection the no-ghost and no-tachyon conditions, and the positivity of central charges  are satisfied only for $C=0$.

\section{Conclusion}
In this paper we have shown that both the TMG  \cite{Deser:1981wh} and the MMG \cite{Bergshoeff:2014pca}
can be treated as spontaneously broken phases of the Chern-Simons theory based on the Hietarinta/Maxwell algebra. In general, the spontaneous symmetry breaking in the HMCSG theory leads to a more general class of minimal massive gravities propagating a single massive spin-2 mode and having two more coupling parameters with respect to the MMG. For a certain range of the parameters these models have neither tachyons nor ghosts and their asymptotic algebra has positive central charges thus giving rise to unitary boundary CFTs. A further more detailed analysis of these models in the AdS/CFT context might be of interest.

 As a generalization of the results of this paper, it would be interesting to identify the group-theoretical structure of Chern-Simons theories whose symmetry breaking gives rise e.g. to ``New", ``General" \cite{Bergshoeff:2009hq, Bergshoeff:2009aq}, ``Zwei-Dreibein" \cite{Bergshoeff:2013xma,Afshar:2014dta} and ``Exotic" Massive Gravities \cite{Ozkan:2018cxj,Ozkan:2019iga,Afshar:2019npk}, for more references see \cite{Merbis:2014vja}. And of course the most challenging issue is to find an Englert–Brout–Higgs–Guralnik–Hagen–Kibble mechanism which might lead to such a symmetry breaking.

 Another interesting direction is to look for a relation of the HMCSG to a ``simple" theory of $3d$ massive gravity constructed and studied in \cite{Geiller:2018ain,Geiller:2019dpc}. The simplicity of this model is due to the fact that it contains only two one-form fields, a dreibein and a would-be Lorentz spin connection, but the local Lorentz symmetry in this model is broken. For a certain choice of the parameters in the letter its field equations reproduce those of the MMG. A question is whether for a more general range of the parameters the simple massive gravity may also reproduce the equations of motion of the HMCSG theory constructed in this paper (upon solving for ${h}^a$ in \eqref{eomnew2}).

It might also be of interest to consider supersymmetric and higher-spin extensions of these models elaborating on the constructions obtained e.g. in  \cite{Fierro:2014lka,Concha:2015woa, Caroca:2017izc,Chernyavsky:2019hyp,Salgado-Rebolledo:2019kft,Caroca:2019dds,Ozdemir:2019tby,Concha:2019mxx,Concha:2020sjt}.

Regarding supersymmetric generalizations, let us make the following final remark. The simplest extension of the Maxwell algebra \eqref{Maxwell} by a two-component Majorana spinor generator $Q_\alpha$ \cite{Soroka:2004fj} is such that $[Q_\alpha,P_a]=[Q_\alpha,Z_a]=0$ and $\{Q_\alpha,Q_\beta\}=2\gamma^{a}_{\alpha\beta}Z_a$, i.e. the anti-commutator of $Q$ can only close on $Z$ due to Jacobi identities. Hence, this simplest super-Maxwell algebra is not an extension of the conventional $N=1, D=3$ super-Poincar\'e algebra. On the contrary, the similar supersymmetric extension of the Hietarinta algebra \eqref{Algebra_Maxwell} is the extension of the simple $3d$ super-Poincar\'e algebra since in this case the Jacobi identities allow $\{Q_\alpha,Q_\beta\}=2\gamma^{a}_{\alpha\beta}P_a$. This gives one more evidence to the fact that the physical models based on the two versions of the same algebra are a priori different.

\section*{Acknowledgements}
The authors are thankful to Eric Bergshoeff, Axel Kleinschmidt,
Wout Merbis and Paul Townsend for interest to this work and useful comments.
DC and NSD are grateful  to INFN, Padova for hospitality and financial support during an intermediate stage of this work. NSD wishes to thank Albert Einstein Institute, Potsdam for hospitality during the final phase of this paper. Work of DC was supported by the Foundation for the Advancement of Theoretical Physics and Mathematics “BASIS”. NSD is partially supported by the Scientific and Technological
Research Council of Turkey (T\"{u}bitak) Grant No.116F137.

\providecommand{\href}[2]{#2}\begingroup\raggedright\endgroup

\end{document}